\documentclass[aps,prb,twocolumn,showpacs]{revtex4}
\usepackage{amssymb}
\usepackage{color}
\usepackage{graphicx}
\usepackage{dcolumn}
\usepackage{bm}

\begin{document}

\title{Coexistence of Itinerant Electrons and Local Moments in Iron-Based
Superconductors}
\author{Su-Peng Kou$^{1}$, Tao Li$^{2}$, and Zheng-Yu Weng$^{3}$}
\email{weng@tsinghua.edu.cn}
\affiliation{$^{1}$Department of Physics, Beijing Normal University, Beijing, 100875,
China\\
$^{2}$Department of Physics, Renmin University of China, Beijing, 100872,
China \\
$^{3}$Center for Advanced Study, Tsinghua University, Beijing, 100084, China}
\date{\today }

\begin{abstract}
In view of the recent experimental facts in the iron-pnictides, we make a
proposal that the itinerant electrons and local moments are simultaneously
present in such multiband materials. We study a minimal model composed of
coupled itinerant electrons and local moments to illustrate how a consistent
explanation of the experimental measurements can be obtained in the leading
order approximation. In this mean-field approach, the spin-density-wave
(SDW) order and superconducting pairing of the itinerant electrons are not
directly driven by the Fermi surface nesting, but are mainly induced by
their coupling to the local moments. The presence of the local moments as
independent degrees of freedom naturally provides strong pairing strength
for superconductivity and also explains the normal-state linear-temperature
magnetic susceptibility above the SDW transition temperature. We show that
this simple model is supported by various anomalous magnetic properties and
isotope effect which are in quantitative agreement with experiments.
\end{abstract}

\maketitle

\section{Introduction}

Since the discovery of superconductivity (SC) in iron pnictides\cite%
{kamihara,wen,chenxh,nlwang,ren}, with $T_{c}$ being quickly raised to $55%
\mathrm{\ K}$\cite{ren}, intensive attentions have been focused on possible
underlying mechanisms. With the\ neutron scattering measurement\cite%
{cruz,McGuire} subsequently establishing the fact that an SDW order exists
in the undoped \textrm{LaOFeAs} compound below $T_{\mathrm{SDW}}\simeq 134$ $%
\mathrm{K}$, which has been later generically found in other iron pnictides%
\cite{huang,ychen,SrFeAs,BaFeAs}, the interplay between SC and
antiferromagnetism (AF) has become a central issue.

The iron 3$d$-electrons are believed quite itinerant with their hybridized
multi-orbitals forming multiple Fermi pockets at the Fermi level\cite{cao}.
Many theoretical efforts\cite{dong,cao,singh,maz.kur,ZDWang,li,Qi,lee,hui}
are based on itinerant approaches in searching for possible SDW and SC
mechanisms responsible for the iron pnictides. This kind of theory is
generally sensitive to the detailed band structure where the Fermi surface
nesting effect is important. As shown by a renormalization group analysis%
\cite{hui}, such an itinerant model does possess the instabilities towards
the SDW and SC orderings. However, how to reach high-$T_{c}$ in the SC phase
and at the same time have a self-consistent description of the magnetic
phase within a unified framework remains a challenge.

Alternatively local moment descriptions have been also promoted\cite%
{Si,weng,cenker-sachdev,jphu,chen,zhang} in view of the $d$-electrons, local
Coulomb and Hund's rule interactions in the iron pnictides, as opposed to
the itinerant RPA-type treatment. Of them the so-called \textrm{J}$_{1}$%
\textrm{-J}$_{2}$ model which emphasizes the As-bridged superexchange
couplings\cite{yildirim,Si,zhong-yi} between the nearest neighboring (NN)
and next nearest neighboring (NNN) local moments of the irons has been used
due to its natural tendency to form the collinear AF order at low
temperature. The local moment approach is especially appealing over the
itinerant one in explaining the anomalous large linear-temperature
susceptibility in the normal state over a wide temperature regime\cite{zhang}%
. However, how this localized spin picture can be meaningfully applied to a
metallic material (albeit a bad metal in the undoped case of the
iron-pnictides) remains unclear. Whether the doping effect is similar to
that of the cuprates as described by a multiband \textrm{t-J}$_{1}$\textrm{-J%
}$_{2}$ like model is also controversial.
\begin{figure}[tp]
\centerline{
    \includegraphics[width=2.5in]{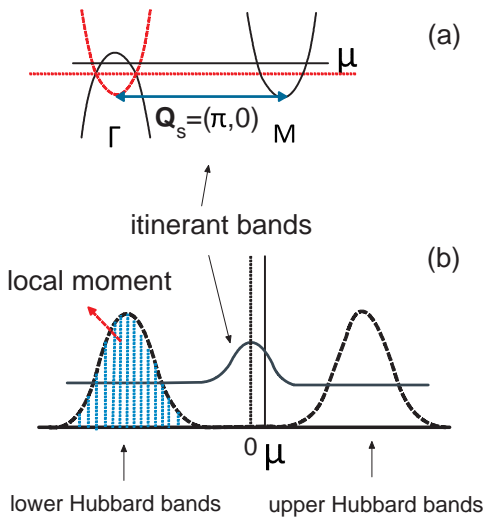}
    } %\vspace{-2cm}
\caption{(color online) A schematic illustration of itinerant electron and
local moment bands and the profiles of density of states in the present
model. (a) A simplified picture of two itinerant bands: A hole-like band
near $\Gamma $ point and an electron-like one near $M$ point, which are
separated by a moment $\mathbf{Q}_{s}=(\protect \pi ,0)$; (b) The density of
states for the itinerant bands shows an enhancement at chemical potential $%
\protect \mu $ near the origin where both the hole and electron pockets
contribute (see (a)); The local moment as an independent degrees of freedom
is contributed by a filled lower Hubbard band, with the Mott gap crossing
the Fermi level such that it does not contribute to the low-energy charge
dynamics. The Hund's rule coupling between the itinerant electrons at the
Fermi level and the local moments will dictate the low-energy physics. }
\label{Fig1}
\end{figure}

In this paper, we point out that if \emph{both} itinerant electrons and
local moments are allowed to simultaneously present in the system, then many
basic properties of the iron pnictides can be naturally accommodated by a
single framework in the leading order approximation. To illustrate the
point, we study a highly simplified model with the local moments and
itinerant electrons coupled together by a Hund's rule coupling as
schematically illustrated in Fig. \ref{Fig1}. We show that an SDW order of
the itinerant electrons can take place simultaneously with a collinear AF
order of the local moments, while the ordering would be absent in either
degrees of freedom if they do not couple, clearly different from the Fermi
surface nesting mechanism or the AF ordering in a \textrm{J}$_{1}$\textrm{-J}%
$_{2}$ model. Furthermore, the superconducting pairing of the itinerant
electrons is also driven by the same coupling with the strength reaching
strong coupling. Such a picture is further supported by a series of magnetic
properties both below and above the SDW ordering temperature $T_{\mathrm{SDW}%
},$ which show good agreement with experiments. It is predicted that in
order to consistently account for $T_{\mathrm{SDW}},$ magnetization, spin
gap, uniform susceptibility, as well as the competition between the AF and
SC phases, the (high-temperature) normal state of local moments should be
close to a critical regime of quantum magnets, which can be tested by a
neutron-scattering experiment.

\section{Model Study}

\subsection{Minimal model}

Our model Hamiltonian is composed of three terms
\begin{equation}
H=H_{\mathrm{it}}+H_{J_{0}}+H_{J_{2}}.  \label{H}
\end{equation}
The first term
\begin{equation}
H_{\mathrm{it}}=\sum_{\mathbf{k},\sigma }\left( \varepsilon _{\mathbf{k}%
}-\mu \right) c_{\mathbf{k}\sigma }^{\dagger }c_{\mathbf{k}\sigma }
\label{hband}
\end{equation}%
describes the itinerant electrons forming the hole and electron pockets near
the Fermi energy as illustrated in Fig. 1(a), which are located at the $%
\Gamma $ point and the $M$ point, respectively, and separated by momenta $%
\mathbf{Q}_{s}=(\pi ,0)$ and $(0,\pi )$ [only the former is shown in Fig.
1(a)] in an extended Brillouin zone (BZ). For simplicity we shall assume the
symmetric dispersions for the hole and electron bands, with $\varepsilon _{%
\mathbf{k}}=-\varepsilon _{\mathbf{k+Q}_{s}}$ such that the hole and
electron pockets are exactly nested at $\varepsilon _{\mathbf{k}}=\mu =0$,
where $\mu $ is the chemical potential. Such nesting of the Fermi pockets
will be lifted as the increase (decrease) of $\mu $, which effectively
controls the electron (hole) doping (the undoped case in iron pnictides may
correspond to some small but finite $\left \vert \mu \right \vert $ here).

In contrast to the conventional itinerant approach where the Coulomb
interaction between the itinerant electrons gets enhanced via the Fermi
surface nesting effect, we shall omit such an interaction. Instead, we
emphasize the importance of the coupling between the itinerant electrons and
some preformed local moments via the second term $H_{J_{0}}$ in (\ref{H}),
by a renormalized Hund's rule coupling $J_{0}$ as follows
\begin{equation}
H_{J_{0}}=-J_{0}\sum_{i}\mathbf{M}_{i}\cdot \mathbf{S}_{i}  \label{hj0}
\end{equation}%
where $\mathbf{S}_{i}$ is the spin operator for the itinerant electrons and $%
\mathbf{M}_{i}$ denotes the local moment at site $i$. In the following we
shall focus on the weak $J_{0}$ case where $\mathbf{M}_{i}$ behaves like an
independent degree of freedom. In the strong coupling limit of $J_{0}$, by
contrast, $\mathbf{S}_{i}$ and $\mathbf{M}_{i}$ should be locked together in
strongly correlated regime as has been previously discussed in Ref. \cite%
{weng}.

An essential assumption of this model will be that besides the itinerant
electrons described above, there are some $d$-electrons sitting \emph{below}
the Fermi energy that can also contribute to the low-energy physics by
forming effective local magnetic moments\cite{weng}. Namely, some of the $d$%
-electron multibands can open up a Mott-Hubbard gap crossing the Fermi
energy, due to the on-site Coulomb interaction and the Hund's rule
ferromagnetic (FM) coupling, with the filled lower Hubbard bands giving rise
to an effective local moment as illustrated in Fig. 1(b). These localized $d$%
-electrons may be different from the itinerant electrons as mainly coming
from more isolated\cite{phillips} $d_{\mathrm{x}^{2}\mathrm{-y}^{2}}$ and $%
d_{\mathrm{z}^{2}}$ orbitals. In the following we shall simply assume their
existence and explore the consequences of it.

The third term in (\ref{H}) describes the predominant interaction between
these local moments by a Heisenberg-like model
\begin{equation}
H_{J_{2}}=J_{2}\sum_{\left \langle ij\right \rangle \in A}\mathbf{M}%
_{i}\cdot \mathbf{M}_{j}+J_{2}\sum_{\left \langle ij\right \rangle \in B}%
\mathbf{M}_{i}\cdot \mathbf{M}_{j}  \label{hj2}
\end{equation}%
where the NNN superexchange coupling \thinspace $J_{2}$ is bridged by the As
ions between the diagonal iron sites, with $A$ and $B$ referring to two
sublattices of the square Fe ion lattice, according to the LDA calculation%
\cite{yildirim,zhong-yi} and analysis\cite{Si}. Note that the NN exchange $%
J_{1}$, bridged by the As ions, can be either AF or FM in nature and much
weaker than $J_{2}$ for the isolated $d_{\mathrm{x}^{2}\mathrm{-y}^{2}}$ and
$d_{\mathrm{z}^{2}}$ orbitals due to the symmetry reason \cite%
{Si,phillips,tesan}. Furthermore, the itinerant electrons can effectively
induce an additional NN\ FM interaction between the local moments, which is
assumed to be predominant (to be consistent with the lattice distortion
induced by the SDW ordering observed in the neutron scattering measurement%
\cite{cruz,huang}, see below). Thus we shall neglect the effect of $J_{1}$
to the leading order approximation. Of course, one may always add such $%
J_{1} $ term as well as the Coulomb interaction between the itinerant
electrons into the above highly simplified model to make it more realistic.
But for the purpose of identifying the most essential components and the
simplicity of the model, we shall focus on the minimal model (\ref{H}) in
the following study.

\subsection{Mean field approximation}

\subsubsection{Effective description of the local moments}

According to (\ref{hj2}), the local moments $\mathbf{M}_{i}$ will
antiferromagnetically fluctuate in each \emph{sublattice} of the iron square
lattice, and thus may be redefined by $\mathbf{M}_{i}\equiv Mp_{i}\mathbf{n}%
_{i}$ with $p_{i}\equiv e^{i\mathbf{Q}_{s}\cdot \mathbf{r}_{i}}$ with $%
\mathbf{Q}_{s}=(\pi ,0)$ and $(0,\pi )$ such that the unit vector $\mathbf{n}%
_{i}$ will fluctuate smoothly in each sublattice. For the iron pnictides,
the presence of the Mott-Hubbard gap is not expected to be very large ($\sim
0.6$ \textrm{eV }as indicated in the optical experiment\cite{optical}). It
is enough to \emph{protect} the local moments from amplitude fluctuations
over a wide temperature regime presumably much higher than the SDW ordering
temperature $T_{\mathrm{SDW}}$ as well as $T_{c}$. But it also means that in
reality the local moment $M$ is not quantized and described by a
Heisenberg-like Hamiltonian (with $S=2$ for instance)$.$

Thus it would be more suitable to use a nonlinear $\sigma $-model\cite%
{chakravarty,sachdev} to characterize the low-energy fluctuations of local
moments in replace of (\ref{hj2}):
\begin{equation}
\mathcal{L}_{J_{2}}=\sum \limits_{a=A,B}\left \{ \frac{1}{2g_{0}}\left[
(\partial _{\tau }\mathbf{n}_{a})^{2}+c^{2}(\nabla _{\mathbf{r}}\mathbf{n}%
_{a})^{2}+i\lambda _{a}(\mathbf{n}_{a}^{2}-1)\right] \right \}  \label{Ls}
\end{equation}%
with $c\simeq 4MJ_{2}$ (with the lattice constant of the Fe square lattice
taking as the unit) and $g_{0}\simeq 16J_{2}$. (Note that $\mathbf{n}_{a}$ ($%
a=A$, $B$) here denotes the unit vector in a given sublattice such that two
\emph{separated} N\'{e}el orders would emerge, if $\lambda _{a}=0$ at $T=0$%
.) Denoting $\mathbf{n}_{0}\equiv \left \langle \mathbf{n}_{a}\right \rangle
,$ the fluctuations of $\delta \mathbf{n}\equiv \mathbf{n}-\mathbf{n}_{0}$
is described by the propagator $D_{0}(\mathbf{q},\tau )=-\left \langle
T_{\tau }\delta \mathbf{n}(\mathbf{q},\tau )\mathbf{\cdot }\delta \mathbf{n}%
(-\mathbf{q},0)\right \rangle $ with\cite{chakravarty,sachdev}

\begin{equation}
D_{0}(\mathbf{q},i\omega _{n})=-\frac{3g_{0}}{\omega _{n}^{2}+\Omega _{%
\mathbf{q}}^{2}}  \label{D}
\end{equation}%
where the spin-wave spectrum
\begin{equation}
\Omega _{\mathbf{q}}=\sqrt{c^{2}\mathbf{q}^{2}+\eta ^{2}}
\end{equation}
and $\eta ^{2}\equiv i\lambda _{a}$ with the subscription $a$ being dropped,
which is determined by the condition $\left \langle \left( \mathbf{n}%
_{a}\right) ^{2}\right \rangle =1$ as
\begin{equation}
\left( \mathbf{n}_{0}\right) ^{2}-\beta ^{-1}\sum_{\omega _{n},\mathbf{q\neq
0}}D_{0}(\mathbf{q},i\omega _{n})=1  \label{eta}
\end{equation}%
with $\beta \equiv 1/k_{\mathrm{B}}T$ and $\omega _{n}=2\pi n\beta $. Hence,
without coupling to the itinerant electrons, $\mathbf{n}_{a}$ ($a=A$, $B$)
do not couple to one another, and the local moment $\mathbf{M}_{i}$ governed
by (\ref{Ls}) will intrinsically fluctuate around the two possible $\mathbf{Q%
}_{s}$'s.

\subsubsection{Mean-field theory}

It is important to note that such fluctuations will strongly couple to
itinerant electrons via $H_{J_{0}}$, for the hole Fermi pockets around the $%
\Gamma $ point and electron Fermi pockets around the $M$ point are
approximately connected by the momentum $\mathbf{Q}_{s}$ at small $\mu ,$ as
shown in Fig. 1(a). In particular, driven by $H_{J_{0}}$, the local moments
and particle-hole pairs can simultaneously condense at a specific wavevector
$\mathbf{Q}_{s}$, giving rise to an AF order at a finite mean-field
temperature $T_{\mathrm{SDW}}$, which can be stabilized presumably by a weak
interlayer coupling.
\begin{figure}[tbp]
\centerline{
    \includegraphics[width=2.8in]{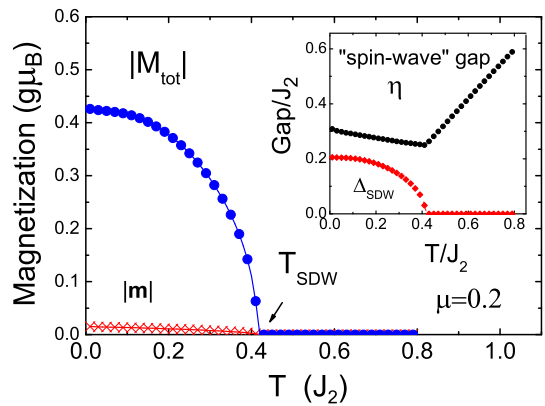}
    }
\caption{(color online) The total magnetization and the induced moment of
itinerant electrons at a fixed chemical potential $\protect \mu =0.2J_{2}$.
Inset: the gaps in the \textquotedblleft spin wave\textquotedblright \
spectrum $\Omega _{\mathbf{q}}$ of local moments and the \textrm{SDW}
spectrum $E_{\mathbf{k}}$ of itinerant electrons. }
\end{figure}

Assume an SDW order parameter for the itinerant electrons
\begin{equation}
\left \langle \mathbf{S}_{i}\right \rangle =\mathbf{m}p_{i}\neq 0
\end{equation}%
with a specific wavevector $\mathbf{Q}_{s}=(\pi ,0)$ in $p_{i}\equiv e^{i%
\mathbf{Q}_{s}\cdot \mathbf{r}_{i}}$. Then a staggered \textquotedblleft
easy-axis field\textquotedblright \ from $H_{J_{0}}$ should be added to the
nonlinear $\sigma $-model in (\ref{Ls}):
\begin{equation}
-J_{0}M\mathbf{m\cdot }\sum \limits_{i}\mathbf{n}_{i}.
\end{equation}%
The resulting Euclidean action is still quadratic in $\mathbf{n}_{i}$ and
can be integrated out in a standard way to give rise to the same expressions
of (\ref{D}) and (\ref{eta}), except that now $\mathbf{n}_{0}$ is determined
by%
\begin{eqnarray}
\mathbf{n}_{0} &\equiv &\left \langle \mathbf{n}_{a}\right \rangle  \nonumber
\\
&=&\left( J_{0}Mg_{0}/\eta ^{2}\right) \mathbf{m}.
\end{eqnarray}%
Thus, no matter how weak $\mathbf{m}$ is, it can always induce a collinear
AF order of the local moments at the same $\mathbf{Q}_{s}$ with $\mathbf{n}%
_{0}\neq 0$. In particular, the independent spin wave spectrum $\Omega _{%
\mathbf{q}}$ of the local moment gains a finite gap $\eta \neq 0$, in
contrast to the limit of $\mathbf{m}=0$ where $\mathbf{n}_{0}\neq 0$ can
only occur at $\eta =0$ in a pure $J_{2}$ nonlinear $\sigma $-model\cite%
{chakravarty,sachdev}.

Self-consistently, with $\mathbf{n}_{0}\neq 0$ a finite $\mathbf{m}$ will
always be induced in the itinerant electrons. At the mean-field level, it is
governed by
\begin{equation}
H_{\mathrm{it}}-J_{0}M\mathbf{n}_{0}\cdot \sum_{i}p_{i}\mathbf{S}_{i},
\end{equation}%
leading to the following band folding and reconstruction
\begin{equation}
\sum \nolimits_{\mathbf{k},\sigma }^{\prime }\left[ \left( -E_{\mathbf{k}%
}-\mu \right) \alpha _{\mathbf{k}\sigma }^{\dagger }\alpha _{\mathbf{k}%
\sigma }+\left( E_{\mathbf{k}}-\mu \right) \beta _{\mathbf{k}\sigma
}^{\dagger }\beta _{\mathbf{k}\sigma }\right]
\end{equation}%
where the summation over $\mathbf{k}$ is restricted within the reduced
magnetic BZ defined by $\mathbf{Q}_{s}$ and the itinerant electron bands are
split into $\alpha $ and $\beta $ bands, with the Bogoliubov transformation%
\begin{eqnarray}
c_{\mathbf{k}\sigma } &=&u_{\mathbf{k}}\alpha _{\mathbf{k}\sigma }-v_{%
\mathbf{k}}\sigma \beta _{\mathbf{k}\sigma },  \nonumber \\
c_{\mathbf{k+Q}_{s}\sigma } &=&v_{\mathbf{k}}\sigma \alpha _{\mathbf{k}%
\sigma }+u_{\mathbf{k}}\beta _{\mathbf{k}\sigma }.
\end{eqnarray}%
Here $u_{\mathbf{k}}=\left[ (1-\epsilon _{\mathbf{k}}/E_{\mathbf{k}})/2%
\right] ^{1/2}$, $v_{\mathbf{k}}=\left[ (1+\epsilon _{\mathbf{k}}/E_{\mathbf{%
k}})/2\right] ^{1/2}$, and the electron excitation spectrum becomes
\begin{equation}
E_{\mathbf{k}}=\sqrt{\epsilon _{\mathbf{k}}^{2}+\Delta _{\mathrm{SDW}}^{2}}
\end{equation}%
in which the SDW gap
\begin{equation}
\Delta _{\mathrm{SDW}}\equiv \frac{J_{0}M}{2}\left \vert \mathbf{n}%
_{0}\right \vert .
\end{equation}%
Finally, the self-consistent mean-field equation reads
\begin{equation}
\left \vert \mathbf{m}\right \vert =\frac{\Delta _{\mathrm{SDW}}}{N}\sum
\nolimits_{\mathbf{k}}^{\prime }\frac{1}{E_{\mathbf{k}}}\left( n_{\mathbf{k}%
\alpha }-n_{\mathbf{k}\beta }\right) ,
\end{equation}%
where $n_{\mathbf{k}\alpha }=1/\left( e^{-\beta (E_{\mathbf{k}}+\mu
)}+1\right) $ and $n_{\mathbf{k}\beta }=1/\left( e^{\beta (E_{\mathbf{k}%
}-\mu )}+1\right) .$

\subsection{Physical properties}

\subsubsection{Mean-field results}

Solving the above mean-field equations, one can determine $T_{\mathrm{SDW}}$
for the collinear AF order and the total magnetization defined by
\begin{equation}
\mathbf{M}_{\mathrm{tot}}\equiv g\mu _{\mathrm{B}}\left( M\mathbf{n}_{0}+%
\mathbf{m}\right) ,
\end{equation}%
$(g=2).$ The results are shown in Fig. 2. Here we have fixed the parameters $%
M=0.8$ and $J_{0}=J_{2}$ throughout the paper, with $J_{2}$ tunable. The
hole dispersion of the itinerant electrons is parameterized as $\varepsilon
_{\mathbf{k}}=-\alpha (\mathbf{k}^{2}-k_{0}^{2})$ with $\alpha =7J_{2}$ and $%
k_{0}=0.1\pi $ based on the ARPES measurement for the so-called
\textquotedblleft 122-type\textquotedblright \  \textrm{BaFe}$_{2}$\textrm{As}%
$_{2}$\cite{feng,liu,Ding,zhou}.

It is noted that the above choice of the parameters is not necessarily
optimized. But such a set of parameters can give rise to a quantitative
account of a series of important experimental results. The calculated $%
\left
\vert \mathbf{M}_{\mathrm{tot}}\right \vert \simeq 0.85$ $\mu _{%
\mathrm{B}}$ at $T=0$ (which is independent of $J_{2}$ as shown in Fig. 2)
is close to $0.87$ $\mu _{\mathrm{B}}$ for \textrm{BaFe}$_{2}$\textrm{As}$%
_{2}$\cite{huang,BaFeAs}. By taking $J_{2}=30$ $\mathrm{meV}$ (incidentally
it is comparable to the LDA estimation\cite{zhong-yi} for \textrm{BaFe}$_{2}$%
\textrm{As}$_{2}$)$\mathrm{,}$ one obtains $T_{\mathrm{SDW}}\simeq
0.415J_{2}=144$ $\mathrm{K}$ (compared to the experimental value $143$ $%
\mathrm{K}$\cite{huang}), and in the inset of Fig. 2, a gap $\eta
=0.31J_{2}=9.3$ \textrm{meV }opened up in the spin wave spectrum of the
local moments is also fairly close to the experimental value $9.8$ \textrm{%
meV}\cite{BaFeAs} (a similar fit for \textrm{SrFe}$_{2}$\textrm{As}$_{2}$%
\cite{SrFeAs} can be also obtained)\textrm{.} Note that a proper cut-off
momentum $\Lambda =0.225\pi $ is taken in solving (\ref{eta}) and its role
in general will be discussed later together with the comparison with other
materials.\textrm{\ }

\subsubsection{Uniform magnetic susceptibility}

Using the same set of parameters, the uniform susceptibility
\begin{equation}
\chi _{u}=\chi _{\mathrm{lo}}+\chi _{\mathrm{it}}
\end{equation}%
is calculated and presented in Fig. 3, which exhibits a pseudogap behavior
below $T_{\mathrm{SDW}}$ and a rough linear-temperature dependence in the
normal state, mainly due to the contribution $\chi _{\mathrm{lo}}$ from the
local moments as shown in the inset. As indicated in the latter, both the
magnitude and slope of $\chi _{\mathrm{lo}}$ are also quantitatively
comparable to the experimental measurements\cite{zhang,xhchen,1111chi}. Here
$\chi _{\mathrm{lo}}$ is given by\cite{sachdev}
\begin{equation}
\chi _{\mathrm{lo}}=\frac{2}{3}\chi _{\perp }\mathbf{n}_{0}^{2}+2\beta
^{-1}\sum_{\omega _{n},\mathbf{q}}\frac{-\omega _{n}^{2}+c^{2}\mathbf{q}%
^{2}+\eta ^{2}}{\left( \omega _{n}^{2}+c^{2}\mathbf{q}^{2}+\eta ^{2}\right)
^{2}}
\end{equation}%
in units of $(g\mu _{\mathrm{B}})^{2}$ where $\chi _{\perp }=1/g_{0}$.
\begin{figure}[tbp]
\centerline{
    \includegraphics[width=2.8in]{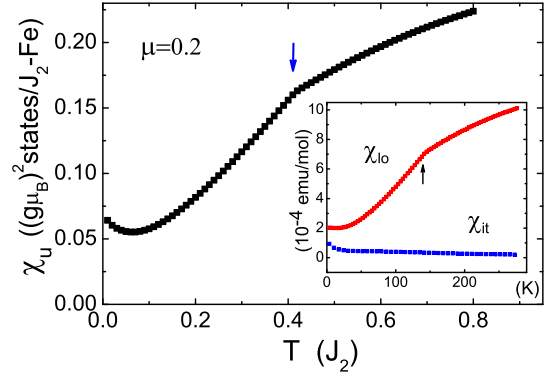}
    }
\caption{(color online) The calculated uniform spin susceptibility $\protect%
\chi _{u}$ shows a linear-T behavior above $T_{\mathrm{SDW}}$ (marked by
arrow) using the same set of parameters as used in Fig. 2. Inset: The
susceptibility $\protect \chi _{\mathrm{lo}}$ contributed by the local
moments and $\protect \chi _{\mathrm{it}}$ from the itinerant electrons are
illustrated separately, in absolute units comparable to the experimental
data with taking $J_{2}=30$ $\mathrm{meV.}$ }
\end{figure}

The contribution from the itinerant electrons is given by%
\begin{equation}
\chi _{\mathrm{it}}=\frac{\beta }{2N}\sum \nolimits_{\mathbf{k}}^{\prime }%
\left[ n_{\mathbf{k}\alpha }\left( 1-n_{\mathbf{k}\alpha })+n_{\mathbf{k}%
\beta }(1-n_{\mathbf{k}\beta }\right) \right]
\end{equation}%
in units of $(g\mu _{\mathrm{B}})^{2}$. $\chi _{\mathrm{it}}$ is mostly
Pauli-like, except for an upturn at low temperature as shown in the inset of
Fig. 3, which is due to the enhanced density of states in the induced SDW
state where the SDW gap is near but not right at the Fermi level. It results
in the upturn of the total susceptibility at low temperature as shown in
Fig. 3.

Thus, the present theory not only gives rise to the large,
linear-temperature dependent susceptibility above $T_{\mathrm{SDW}}$, but
also naturally explains the quick drop of the susceptibility below $T_{%
\mathrm{SDW}}$ due to the gap opening and its upturn at even lower
temperatures generically found in the iron pnictides\cite{xhchen,1111chi}.

\begin{figure}[tbp]
\centerline{
    \includegraphics[width=3.2in]{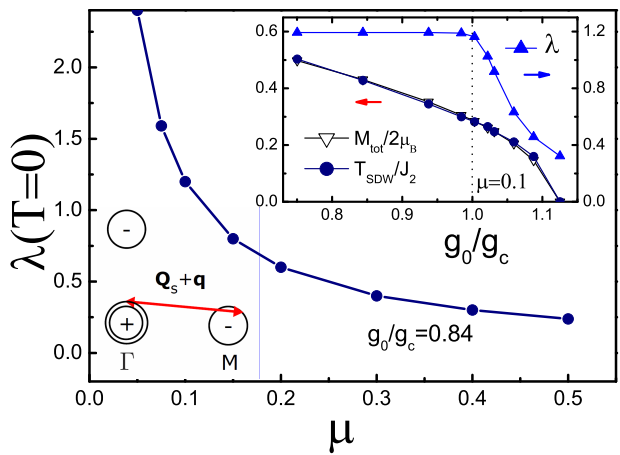}
    }
\caption{(color online) The pairing strength $\protect \lambda $ vs. $\protect%
\mu ,$ with the relative sign change of the pair amplitude shown at the left
bottom. Inset: Variations of $\protect \lambda ,$ $M_{\mathrm{tot}}$, and $T_{%
\mathrm{SDW}}$ vs. the parameter $g_{0}/g_{c}$ of the nonlinear $\protect%
\sigma $-model (\protect \ref{Ls}) (see text). }
\end{figure}

\subsubsection{Pairing strength}

The itinerant electrons can also exchange the quantum fluctuations of the
local moments to form Cooper pairs, which resonantly hop between the
electron and hole pockets with large momentum transfers around $\mathbf{Q}%
_{s}$. The pairing amplitude is s-wave within each Fermi pocket, but has to
change\emph{\ sign} between the two pockets due to exchanging the spin
fluctuation $D_{0}$ [(\ref{D})], consistent with other approaches\cite%
{maz,kur,jphu,hui} and measurement\cite{Ding}.

In Fig. 4, an effective dimensionless pairing strength $\lambda $ as
function of $\mu $ is calculated based on the same set of parameters with
fixing $\mathbf{m}=0,$\thinspace i.e, in the absence of the SDW order. Here $%
\lambda $ is given by
\begin{equation}
\lambda \equiv 2\left( J_{0}M\right) ^{2}N_{F}\left \langle (-1)D_{0}(%
\mathbf{k-k}^{\prime }\mathbf{-Q}_{s},\omega =0)\right \rangle _{\mathrm{FS}}
\label{lmbda}
\end{equation}%
where $N_{F}=1/\left( 4\pi \alpha \right) $ denotes the density of states at
the Fermi energy and the average is over the Fermi pockets. Note that to
properly estimate the strength we have considered two hole pockets at $%
\Gamma $ and two electron pocket at two $M$'s in consistency with ARPES
experiments\cite{Ding}, as marked in the left bottom in Fig. 4, where the
sign change of the pairing amplitude is also indicated. The main panel of
Fig. 4 shows that $\lambda $ is fairly large in a wide regime (it even
diverges at $\mu =0$ due to the artificial nesting effect of the Fermi
pockets). Thus the superconducting phase is expected to strongly compete
with the SDW state at low doping.

We further find that $\mathbf{M}_{\mathrm{tot}}$ and $T_{\mathrm{SDW}}$ are
quite flat as function of $\mu $, insensitive to doping. But the SDW order
is very sensitive to the aforementioned momentum cut-off $\Lambda $, which
decides the critical coupling constant\cite{chakravarty} $g_{c}=4\pi
c/3\Lambda $ in the nonlinear $\sigma $-model (\ref{Ls}). Indeed, as given
in the inset of Fig. 4, both $\mathbf{M}_{\mathrm{tot}}\,$and $T_{\mathrm{SDW%
}}$ are well scaled together and monotonically decrease with the increase of
$g_{0}/g_{c}\propto \Lambda $, whereas $\lambda $ remains flat until
reaching beyond $g_{0}/g_{c}=1$ (on the right hand side of the dashed
vertical line). A quantitative check reveals that a good agreement with
experiments always occurs in the regime that the coupling constant $g_{0}$
is not far from the quantum critical point $g_{c}$. For instance, by
increasing $\Lambda $, say, to $g_{0}/g_{c}\simeq 1.03$, $\left \vert
\mathbf{M}_{\mathrm{tot}}\right \vert $ reduces to $0.42$ $\mu _{\mathrm{B}}$
and $T_{\mathrm{SDW}}\simeq 0.2J_{2}=139$ \textrm{K }by choosing $J_{2}=60$
\textrm{meV}$.$ Then one can get a good quantitative account for the similar
properties of the so-called \textquotedblleft 1111\textquotedblright \
compounds\cite{cruz,ychen},\textrm{\ }like \textrm{LaOFeAs}$.$ Note that as
compared to the \textquotedblleft 122\textquotedblright \ materials the $%
J_{2} $ value here is doubled, which is also comparable to the LDA estimation%
\cite{zhong-yi} for \textrm{LaOFeAs}. Using the BCS formula
\begin{equation}
T_{c}\simeq \omega _{0}\exp \left[ -\left( 1+\lambda \right) /\lambda \right]
,
\end{equation}%
it explains why $T_{c}$ can be much higher in the \textquotedblleft
1111\textquotedblright \ compounds\cite{ren} as $\omega _{0}\propto J_{2}$.

Furthermore, since $\omega _{0}\propto J_{2}$, one can estimate the isotope
effect of the iron mass based on
\begin{equation}
J_{2}\propto 1+\left \langle \left \vert u\right \vert ^{2}\right \rangle
\end{equation}%
(using the fact that the effective\ hopping integral $\propto 1+u$ under a
relative lattice displacement $u$) and $\left \langle \left \vert u\right \vert
^{2}\right \rangle \propto 1/\left( M_{\mathrm{Fe}}\Theta _{D}\right) \propto
$ $\left( M_{\mathrm{Fe}}\right) ^{-1/2}$ as in the Debye-Waller factor ($%
\Theta _{D}$ is the Debye frequency and $M_{\mathrm{Fe}}$ the iron mass),
with
\begin{equation}
\alpha _{\mathrm{SC}}\equiv -d\ln T_{c}/d\ln M_{\mathrm{Fe}}\simeq
0.5[\left \langle \left \vert u\right \vert ^{2}\right \rangle /(1+\left \langle
\left \vert u\right \vert ^{2}\right \rangle )].
\end{equation}
According to the inset of Fig. 4, $T_{\mathrm{SDW}}\propto J_{2}$ and thus
the magnetic isotope coefficient $\alpha _{\mathrm{SDW}}\equiv -d\ln T_{%
\mathrm{SDW}}/d\ln M_{\mathrm{Fe}}$ is related to $\alpha _{\mathrm{SC}}$ by
\begin{equation}
\alpha _{\mathrm{SDW}}\simeq \alpha _{\mathrm{SC}}
\end{equation}%
which is consistent with the recent experimental result\cite{chen-iso}$,$
but the positive $\alpha _{\mathrm{SC}}$ here is in contrast with another
recent experiment\cite{shirage-iso} where a negative $\alpha _{\mathrm{SC}}$
is obtained for samples of high-pressure synthesis. Whether the isotope
effect is positive or negative needs a further experimental clarification.

\section{Conclusion and Discussion}

In this paper, we have proposed a minimal model based on some basic
band-structure and experimental facts, and shown that it can provide a
systematic and quantitative account for a series of anomalous magnetic and
SC properties observed in the iron pnictides.

This is a multiband model, composed of two independent and distinct
components, i.e., the itinerant electrons and local moments, respectively.
Here the charge carriers of the itinerant bands presumably come from the
more extended $d$-orbitals, like $d_{\mathrm{xy}},$ $d_{\mathrm{zx}}$, and $%
d_{\mathrm{yz}}$. The local moments are contributed by the electrons from
some more localized orbitals, say,\ $d_{\mathrm{x}^{2}\mathrm{-y}^{2}}$ and $%
d_{\mathrm{z}^{2}}$ orbitals, forming at much higher temperature than that
of the SDW ordering observed in the experiment. Such local moments are
protected by a Mott gap which is supposed to cross the Fermi level all the
time such that they do not directly contribute to the charge dynamics.
Finally the two independent degrees of freedom, the itinerant electrons and
local moments, are coupled together by the local Hund's rule interaction
inside each iron.

This is a highly simplified model. We have totally omitted the Coulomb
interaction between the itinerant electrons such that the SDW ordering of
the itinerant electrons is not driven by a conventional picture of Fermi
surface nesting effect. For the local moments from the $d_{\mathrm{x}^{2}%
\mathrm{-y}^{2}}$ and/or $d_{\mathrm{z}^{2}}$ orbitals, we have only kept
the dominant NNN \textrm{J}$_{2}$ superexchange coupling, bridged by the $%
\mathrm{As}$ ions, and neglected the NN coupling \textrm{J}$_{1}$ due to the
symmetry reason. So the local moments alone do not form the proper collinear
AF order as seen in experiment, and are effectively described by the
familiar NNN AF nonlinear $\sigma $-models at two sublattices, respectively.

Then we have shown that the magnetic ordering can be simply realized due to
the strong Hund's rule coupling between the two components. Namely the
itinerant electrons can form an SDW order simultaneously with the collinear
AF order of the local moments at the same momentum $\mathbf{Q}_{s}$ below a
mean-field transition temperature $T_{\mathrm{SDW}}$, with $H_{J_{0}}$
gaining a mean-field energy.

Of course, one may further consider the Coulomb interaction in the itinerant
bands which may result in an SDW instability without involving the local
moments. Or one can further introduce the NN coupling \textrm{J}$_{1}$ for
the local moments and obtain a collinear AF state independent of the
itinerant electrons. But the key assumption in this approach is that these
effects are negligible, to the leading order approximation, as compared to
the Hund's rule coupling term and thus are omitted for simplicity. One can
always make the model more realistic by adding more perturbative terms
later. As a matter of fact, in the experimentally observed lattice distortion%
\cite{cruz,huang} accompanying the collinear AF order, the NN spins along
the shorter lattice constant direction are always FM parallel while the
antiparallel NN spins usually correspond to the longer lattice constant\cite%
{huang}, in contrast to a \textrm{J}$_{1}$-driven mechanism which would
prefer an opposite lattice distortion\cite{cenker-sachdev}. This provides a
further support for the mechanism of the Hund's rule coupling, with the FM
spins at the shorter NN sites in favor of the kinetic energy of the
itinerant electrons.

Therefore, it is the Hund's rule coupling term (\ref{hj0}) that makes the
present minimal model nontrivial. By locking the SDW order of the itinerant
electrons and local moments together at the wavevector $\mathbf{Q}_{s}$, an
imperfect \textquotedblleft nesting\textquotedblright \ of the hole and
electron pockets by $\mathbf{Q}_{s}$ and the short-range AF ordering of the
local moments at $\mathbf{Q}_{s}$ are synchronized to optimize the total
energy. One finds that $T_{\mathrm{SDW}},$ the magnitude of the total
magnetization moment, the small spin gap due to the out-of-phase relative
fluctuations of the SDW moments between the itinerant electrons and local
moments, can be quantitatively determined, in good agreement with the
experimental results.

Compared to the conventional Fermi surface nesting mechanism for the
itinerant electrons, the present model naturally predicts the presence of
pre-formed magnetic moments above the SDW ordering temperature. The
calculated normal-state magnetic susceptibility above $T_{\mathrm{SDW}}$
shows a linear-temperature behavior consistent with the experiment in
magnitude and slope, under the same set of parameters, which is difficult to
understand by the itinerant electrons alone. Furthermore, while the Fermi
surface nesting mechanism is difficult to explain why $T_{c}$ is so high in
the iron pnictides, the computed pairing strength between the itinerant
electrons via exchanging the AF fluctuations of the local moments is found
to be easily in strong coupling regime in the present work, using the same
parameters.

On the other hand, as compared to a conventional \textrm{J}$_{1}$\textrm{-J}$%
_{2}$ model where the collinear AF transition can occur alone at low
temperature, in the present model the itinerant electrons play a crucial
role in driving the magnetic ordering by coupling to the local moments, and
by doing so drastically change their own dynamics below $T_{\mathrm{SDW}}$,
resulting in peculiar gap behaviors in static and dynamic susceptibilities
as well as the scattering rate change in optical conductivity, which have
been observed experimentally and are hard to understand if the magnetic
ordering is due to the \textrm{J}$_{1}$\textrm{-J}$_{2}$ superexchange
interaction alone.

As a matter of fact, in order to provide a consistent explanation of some
generic phenomena found in the iron pnictides including the SDW and SC
orders within a single framework, the present model predicts that in the
normal state the local moments should be close to a critical regime of
quantum magnets with the dominant \textrm{J}$_{2}$-type superexchange
interaction, which can be critically tested by a neutron-scattering
experiment. The magnetic state of the local moments is also expected to play
an important role in the superconducting state. Here the reason behind the
rise of $T_{c}$ by doping or pressure in experiment may be not due to a pure
increase of charge carrier number like in the cuprates, but rather due to
the suppression of the collinear AF order in favor of superconductivity, as
the result of the accompanying change in the ratio $g_{0}$/$g_{c}$ near the
quantum critical regime\cite{chakravarty,sachdev1}, as discussed above.
However, how this mechanism can be realized microscopically is beyond the
scope of the present work, which may involve detailed and realistic local
interactions (including $J_{1})$, and will be left for a future study.

\begin{acknowledgments}
We would like to acknowledge stimulating discussions with D.H. Lee, Z.Y. Lu,
Q.H. Wang, Y.Y. Wang, M.Q. Weng, M.W. Wu, T. Xiang, J. Zaanen, G.M. Zhang,
and also thank W. Bao, X.H. Chen, P. C. Dai, H. Ding, D.L. Feng, N.L. Wang,
H.H. Wen, and X. J. Zhou for generously sharing their experimental results.
This work is supported by NSFC, NBRPC, and NCET grants.
\end{acknowledgments}


\begin{thebibliography}{99}
\bibitem{kamihara} Y. Kamihara, et al., J. Am. Chem. Sco. \textbf{128},
10012 (2006); Y. Kamihara, \emph{et al.}, J. Am. Chem. Sco. \textbf{130},
3296 (2008).

\bibitem{wen} H.-H. Wen, \emph{et al}., Europhys. Lett. \textbf{82}, 17009
(2008).

\bibitem{chenxh} X. H. Chen, \emph{et al.}, Nature (London) \textbf{453},
761 (2008).

\bibitem{nlwang} G. F. Chen, \emph{et al.}, Phys. Rev. Lett. \textbf{100},
247002 (2008).

\bibitem{ren} Z.A. Ren, \emph{et al.}, Europhys. Lett. \textbf{83}, 17002
(2008).

\bibitem{cruz} C. de la Cruz, \emph{et al.}, Nature (London) \textbf{453},
899 (2008).

\bibitem{McGuire} M. A. McGuire, \emph{et al.}, Phys. Rev B \textbf{78},
094517 (2008).

\bibitem{huang} Q. Huang, \emph{et al.}, Phys. Rev. Lett. \textbf{101},
257003 (2008).

\bibitem{ychen} Y. Chen, \emph{et al., }Phys. Rev. B \textbf{78}, 064515
(2008)

\bibitem{SrFeAs} J. Zhao, \emph{et al}., Phys. Rev. Lett. \textbf{101},
167203 (2008).

\bibitem{BaFeAs} K. Matan, \emph{et al.,} Phys. Rev. B\textbf{\ 79}, 054526
(2009).

\bibitem{dong} J. Dong, \emph{et al.}, Europhys. Lett. \textbf{83}, 27006
(2008).

\bibitem{cao} C. Cao, \emph{et al.,} Phys. Rev. B \textbf{77}, 220506 (R)
(2008).

\bibitem{singh} D. J. Singh and M. H. Du, Phys. Rev. Lett. \textbf{100},
237003 (2008).

\bibitem{maz} I. Mazin, \emph{et al.}, Phys. Rev. Lett. \textbf{101}, 057003
(2008).

\bibitem{kur} K. Kuroki, \emph{et al.}, Phys. Rev. Lett. \textbf{101},
087004 (2008).

\bibitem{ZDWang} Q. Han, \emph{et al.}, Europhys. Lett. \textbf{82}, 37007
(2008); Z. J. Yao, \emph{et al.,} New J. Phys. \textbf{11}, 025009 (2009).

\bibitem{li} T. Li, \emph{et al., }J. Phys: Condens. Matter \textbf{20},
425203 (2008).

\bibitem{Qi} S. Raghu, \emph{et al., }Phys. Rev. B \textbf{77}, 220503
(2008); X.L. Qi, \emph{et al.,} arXiv:0804.4332 (unpublished).

\bibitem{lee} P. A. Lee and X. G. Wen, Phys. Rev. B \textbf{78}, 144517
(2008).

\bibitem{hui} F. Wang, \emph{et al.}, Phys. Rev. Lett. \textbf{102}, 047005
(2009); H. Zhai, \emph{et al., }Europhys. Lett. \textbf{85}, 37005 (2009).

\bibitem{Si} Q. Si and E. Abrahams, Phys. Rev. Lett. \textbf{101}, 076401
(2008).

\bibitem{weng} Z. Y. Weng, arXiv:0804.3228.

\bibitem{cenker-sachdev} C. Xu, \emph{et al.}, Phys. Rev. B \textbf{78},
020501 (R) (2008).

\bibitem{jphu} C. Fang, \emph{et al.}, Phys. Rev. B \textbf{77}, 224509
(2008); K. Seo, \emph{et al.}, Phys. Rev. Lett. \textbf{101}, 206404 (2008).

\bibitem{chen} W. Q. Chen, \emph{et al., }Phys. Rev. Lett. \textbf{102},
047006 (2009).

\bibitem{zhang} G. M. Zhang, \emph{et al.}, arXiv:0809.3874 (unpublished).

\bibitem{yildirim} T. Yildirim, Phys. Rev. Lett. \textbf{101}, 057010 (2008).

\bibitem{zhong-yi} F. Ma, \emph{et al.}, Phys. Rev. \textbf{78}, 224517
(2008).

\bibitem{optical} W. Z. Hu, \emph{et al., }Phys. Rev. Lett. \textbf{101},
257005 (2008).

\bibitem{phillips} J. Wu, \emph{et al.}, Phys. Rev. Lett. \textbf{101},
126401 (2008).

\bibitem{tesan} V. Cvetkovic and Z. Tesanovic, Europhys. Lett. \textbf{85},
37002 (2009).

\bibitem{chakravarty} S. Chakravarty, \emph{et al.}, Phys. Rev. B \textbf{39}%
, 2344 (1989).

\bibitem{sachdev} S. Sachdev, \emph{Quantum Phase Transitions,} Cambridge
University Press (1999).

\bibitem{feng} L. X. Yang, \emph{et al.}, Phys. Rev. Lett. \textbf{102},
107002 (2009).

\bibitem{liu} C. Liu, \emph{et al.,} Phys. Rev. Lett. \textbf{101}, 177005
(2008).

\bibitem{Ding} H. Ding \emph{et al., }Europhys. Lett. \textbf{83}, 47001
(2008).

\bibitem{zhou} H. Liu, \emph{et al.}, Phys. Rev. B \textbf{78}, 184514
(2008).

\bibitem{xhchen} X. F. Wang, \emph{et al.}, Phys. Rev. Lett. \textbf{102},
117005 (2009).

\bibitem{1111chi} R. Klingeler, \emph{et al.}, arXiv:0808.0708 (unpublished).

\bibitem{sachdev1} S. Sachdev, \emph{et al.}, Phys. Rev. B \textbf{51},
14874 (1995).

\bibitem{chen-iso} R. H. Liu, \emph{et al.}, arXiv:0810.2694.

\bibitem{shirage-iso} P. M. Shirage, \emph{et al.}, arXiv:0903.3515
(unpublished).
\end{thebibliography}
\end{document}